**Accelerated quantum control using superadiabatic dynamics in a solid-state lambda system**

Brian B. Zhou[1], Alexandre Baksic[2], Hugo Ribeiro[2], Christopher G. Yale[1], F. Joseph Heremans[1,3], Paul C. Jerger[1], Adrian Auer[4], Guido Burkard[4], Aashish A. Clerk[2], David D. Awschalom[1,3†]

[1] *Institute for Molecular Engineering, University of Chicago, Chicago, Illinois 60637, USA*

[2] *Department of Physics, McGill University, Montreal, Quebec H3A 2T8, Canada*

[3] *Materials Science Division, Argonne National Laboratory, Argonne, Illinois 60439, USA*

[4] *Department of Physics, University of Konstanz, D-78457 Konstanz, Germany*

† *Correspondence to: awsch@uchicago.edu*

**Adiabatic evolutions find widespread utility in applications to quantum state engineering[1], geometric quantum computation[2], and quantum simulation[3]. Although offering robustness to experimental imperfections, adiabatic processes are susceptible to decoherence due to their long evolution time. A general strategy termed 'shortcuts to adiabaticity'[4–9] (STA) aims to remedy this vulnerability by designing fast dynamics to reproduce the results of slow, adiabatic evolutions. Here, we implement a novel STA technique known as 'superadiabatic transitionless driving'[10] (SATD) to speed up stimulated Raman adiabatic passage[1,11–14] (STIRAP) in a solid-state lambda ($\Lambda$) system. Utilizing optical transitions to a dissipative excited state in the nitrogen-vacancy (NV) center in diamond, we demonstrate the accelerated performance of different shortcut trajectories for population transfer and for the initialization and transfer of coherent superpositions. We reveal that SATD protocols exhibit robustness to dissipation and experimental uncertainty, and can be optimized when these effects are present. These results motivate STA as a promising tool for controlling open quantum systems comprising individual or hybrid nanomechanical, superconducting, and photonic elements in the solid state[11–16].**

Coherent control of quantum states is a common building block behind quantum technologies for sensing, information processing, and simulation. A powerful class of such techniques is based on the adiabatic theorem, which assures that a system will remain in the same instantaneous eigenstate if changes to the system are sufficiently slow. While adiabatic techniques are attractive for their robustness to experimental fluctuations, their effectiveness is limited when decoherence occurs on timescales comparable to that required by the adiabatic theorem. To mitigate this drawback, exact dynamics resulting from specially-designed control fields were proposed to realize the same purpose as adiabatic evolutions do, but without condition on the evolution time[8]. These approaches for accelerating adiabatic protocols are collectively known as 'shortcuts to adiabaticity' (STA)[4–10]. Beyond providing practical benefits, they address quantum mechanical limits on the speed of dynamical evolution and the efficiency of thermodynamics[8].

Among the strategies for STA is counterdiabatic (or transitionless) driving, which introduces, in its simplest formulation, an auxiliary control field that precisely cancels nonadiabatic transitions between the adiabatic (instantaneous) eigenstates of an initial Hamiltonian. Offering broad applicability, counterdiabatic driving has been demonstrated to speed up state transfer in two-level quantum systems[17,18], as well as the expansion[19] and transport[20] of trapped atoms. Theoretically, it has also been proposed to facilitate the preparation of many-body states for quantum simulation[21]. However, implementation of the counterdiabatic field, particularly in higher-dimensional systems, can be challenging as it may require complex experimental resources to realize interactions absent in the original Hamiltonian. Moreover, as STA protocols generally assume ideal (unitary) evolution and perfect implementation, their robustness to dissipation and experimental uncertainty remains an open question.

To explore these issues, we demonstrate a generalization[10,22] of the counterdiabatic strategy to expedite coherent manipulations in a three-level Λ system. Our starting point is STIRAP, whereby population transfer between two levels is mediated by their coupling to a third intermediate level (Fig. 1a). An overlapping sequence of two driving fields, with the Stokes pulse $\Omega_S(t)$ preceding the pump pulse $\Omega_P(t)$, guides the system along a dark state that evolves from the initial to target state without occupying the intermediate level[1]. For STIRAP, however, achieving transitionless evolution in the adiabatic basis requires a counterdiabatic field that directly couples the initial and target levels[6], a previously unnecessary interaction. To maintain the full utility of STIRAP by introducing only modifications to the original Stokes and pump fields, we instead enforce transitionless evolution in a dressed state basis that reproduces the desired initial and final conditions, but does not track the adiabatic evolution (see Ref. 10 and Supplementary Section 1). This novel approach, which we term 'superadiabatic' transitionless driving (SATD), is illustrated in Fig. 1b. An example superadiabatic shortcut (solid, red line) drives the same transfer as the adiabatic evolution (dashed, red line) does, but via an alternate trajectory, which defines the 'dressed dark state'.

In contrast to the adiabatic evolution for STIRAP, our shortcut trajectories deliberately occupy the intermediate state and hence are sensitive to its dissipation. However, the degree of occupation can be tailored by choice of the dressed dark state[10]. For the case of resonant STIRAP (one and two-photon detunings $\Delta = 0$, $\delta = 0$, respectively, as labeled in Fig. 1a), we start with a pulse (Vitanov shape) known to be adiabatically optimal[23] and display how it evolves to ensure finite-time, transitionless driving with respect to two distinct choices for the dressed basis: the 'superadiabatic' basis[22] (SATD protocol) and a modified basis (MOD-SATD protocol). The latter is derived from the former by reducing the intermediate level occupation (Supplementary Section 1). The form of the superadiabatic $\Omega_S(t)$ (shown in Fig. 1c) and $\Omega_P(t)$ pulses are determined by the shape parameter $A_{shape} = \Omega_{shape}/\Omega_{min}$, where $\Omega_{shape}$ denotes the maximum Rabi coupling assumed in the theoretical pulse calculation and $\Omega_{min}$ is a reference value proportional to the inverse of the pulse duration $L$ (see Methods). Under unitary evolution, perfect state transfer is achieved by employing pulses with $A_{shape}$ matching the experimental adiabaticity $A = \Omega/\Omega_{min}$, where $\Omega$ denotes the actual value of the experimental Rabi coupling. As $A \to \infty$, conditions are fully adiabatic and the superadiabatic correction vanishes to reproduce the original Vitanov shape ($A_{shape} \to \infty$). Alternatively, $A = 1$ corresponds to the most non-adiabatic condition ($\Omega = \Omega_{min}$) whose corrected pulse ($A_{shape} = 1$) does not exceed the original maximum Rabi coupling.

We realize our protocols using optical driving in a solid-state $\Lambda$ system hosted by a single NV center in diamond at low temperature (*T* = 5.5 K). With its rich energy level structure, spin-photon interface, and natural coupling to proximal nuclear spins, this defect spin presents a dynamic arena for techniques in quantum information[12,24–28] and metrology[29]. Passing a single tunable laser (637.2 nm) through a phase electro-optic modulator (PEOM) produces frequency harmonics to resonantly excite both the $|-1\rangle$ and $|+1\rangle$ ground state spin levels, Zeeman-split by 1.414 GHz, to the $|A_2\rangle$ spin-orbit excited state, which serves as the intermediate state for STIRAP (Fig. 1a and 1d). The intensities of the harmonics are subsequently modulated by an amplitude electro-optic modulator (AEOM), such that coordinated control of the PEOM and AEOM with a 10 GHz arbitrary waveform generator produces the temporal profiles for $\Omega_S(t)$ and $\Omega_P(t)$ used in superadiabatic driving (Fig. 1d and Methods).

As we incorporate the excited state $|A_2\rangle$ into our shortcut dynamics, its spontaneous emission lifetime $T_1$ and orbital dephasing rate $\Gamma_{orb}$ provide us unique insight on the effect of dissipation on STA. Moreover, spectral diffusion of the excited level, a ubiquitous feature of solid state systems, probes the robustness of our protocol to fluctuations from one-photon resonance. We first illustrate these effects through measurement of the photoluminescence (PL) during constant excitation of the $|-1\rangle$ to $|A_2\rangle$ transition (Fig. 1e). Proportional to the occupation of $|A_2\rangle$, the PL reveals coherent Rabi oscillations between the ground and excited states[30]. These oscillations damp due to a combination of spectral diffusion (estimated as Gaussian-distributed with standard deviation $\sigma \sim 31$ MHz), lifetime $T_1$ = 11.1 ns, and dephasing $\Gamma_{orb} = 1/(18\text{ ns})$ for the NV center used (parameter determination is described in Supplementary Sections 2 and 3). Moreover, an overall decay stems predominantly from trapping into $|+1\rangle$, the dark state defined by our driving field[24,31,32].

We begin by examining the effectiveness of our superadiabatic protocols as a function of the maximum optical Rabi strength $\Omega$ of the Stokes and pump pulses. For a constant pulse duration $L = 16.8$ ns (with an additional 2 ns buffer at each end for switching on and off the optical fields), the weakest Rabi coupling that can be corrected without exceeding the maximum amplitude of the adiabatic pulse is $\Omega_{min} = 2\pi \cdot 72.6$ MHz ($\propto L^{-1}$, see Methods). After initializing into $|-1\rangle$, we transfer the population into $|+1\rangle$ using STIRAP pulses of varying $\Omega$ to explore different regimes of the experimental adiabaticity $A = \Omega/\Omega_{min}$. In Fig. 2a, we demonstrate that SATD and MOD-SATD pulses with shape parameter $A_{shape} = A$, as prescribed by theory for unitary evolution, significantly outperform the Vitanov (adiabatic) shape in transfer efficiency. Despite the presence of dissipation and spectral instability, which preclude the perfect efficiencies predicted in their absence, the superadiabatic protocols realize enhancements of >40% in absolute efficiency over the adiabatic protocol as conditions become increasingly non-adiabatic ($A \to 1$). This indicates the relative importance of minimizing transitions out of each protocol's dark state. Furthermore, the design of MOD-SATD to reduce the excited state occupation in the evolution of the dressed dark state decreases its exposure to dissipation and allows it to surpass SATD in efficiency (Fig. 2a).

To investigate the robustness of these protocols, we implement pulse shapes deviating from $A_{shape} = A$, anticipating applications where errors in the determination of $A$ or pulse shape may occur. In Fig. 2b, we fix both $\Omega$ and $L$, resulting in $A = 1.58$, and then apply pulses with $A_{shape}$ ranging from 1 to 4. We find that both SATD and MOD-SATD achieve better transfer efficiency than the adiabatic protocol (magenta bar) for a wide range of pulse shapes. Moreover, within each family of modified shapes, the pulse shape that maximizes the transfer efficiency does not correspond to $A_{shape} = A$, as expected in the absence of dissipation, but to a value $A_{shape}^{opt} < A$. In Fig. 2c (left), we confirm this trend as $A$ varies via the optical power. A linear fit (cyan line) to the extracted transfer efficiency maxima (black points) yields $A_{shape}^{opt} = 0.81(2)\, A + 0.09(3)$ for SATD (see Supplementary Section 4.2 for MOD-SATD).

While part of the deviation from $A_{shape}^{opt} = A$ likely results from attenuation of the pulse shape in the experimental hardware, our master equation model produces a similar deviation simply by incorporating the measured lifetime ($T_1$), dephasing ($\Gamma_{orb}$), and spectral diffusion of the $|A_2\rangle$ excited state (Fig. 2c, right). Physically, the presence of the dissipative mechanisms and fluctuations from one-photon resonance damp transitions to and from the intermediate level, requiring more accentuated drive pulses ($A_{shape}^{opt} < A$) to mimic the optimal trajectory found in the unitary and zero-detuning ($\Delta = 0$) limit. In Supplementary Section 4.3, we present data using deliberate off-resonant driving that support a shift toward more accentuated optimal pulses for nonzero detuning, as similarly induced by spectral diffusion. Taking a wider perspective, the broad funnel of enhanced transfer efficiency in Fig. 2c demonstrates that these protocols are resilient to moderate dissipation and to potential imperfections in real applications, such as in the pulse shape ($A_{shape}$), laser intensity ($\Omega$), or laser frequency ($\Delta$).

In Fig. 2d, we confirm the dynamics of our superadiabatic shortcuts by measuring the time-resolved PL during the adiabatic pulse and during the optimal SATD and MOD-SATD pulses for $\Omega = 2\pi \cdot 113$ MHz. Strikingly, the converted $|A_2\rangle$ populations peak near the center of the pulse sequence for the shortcut protocols, prior to when they peak for the adiabatic protocol.

This offset is a signature of the shortcut's aim to preemptively place population into the intermediate state during the first half of the sequence and to coherently retrieve that population during the second half (see simulations in Supplementary Section 4.4). In contrast, any population in $|A_2\rangle$ during the adiabatic pulse is unintentional and detrimental to fidelity. Moreover, we verify that the maximal $|A_2\rangle$ population for MOD-SATD is ~20% lower (relatively) than SATD, consistent with its theoretical design[10]. Some parasitic $|A_2\rangle$ population during the shortcuts, such as the weak second bump in the SATD trace, is apparent due to the imperfect initialization and fidelity of our implementation.

To characterize the speed-up of our superadiabatic shortcuts, we turn to measurements of the transfer efficiency by varying the adiabaticity $A$ through the pulse length $L$ (i.e., $\Omega_{min} \propto L^{-1}$). As shown in Fig. 3 for constant $\Omega = 2\pi \cdot 122$ MHz, the optimal SATD and MOD-SATD pulses maintain much higher transfer efficiencies as $L$ is reduced. Interpolating between the data points, we infer that the pulse length $L$ for MOD-SATD (SATD) required to reach a transfer efficiency of 90% is ~2.7 (2.0) times shorter than that for the adiabatic pulse (Fig. 3 inset). For the coupling strength shown, our shortest superadiabatic protocol length of 12.6 ns, which maintains efficiencies >85%, is just over twice the quantum speed-limit (QSL) for transfer between two levels through an intermediate state: $L_{QSL} = \sqrt{2}\pi/\Omega$ = 5.8 ns. This QSL transfer utilizes a 'hybrid' rectangular pulse scheme that significantly occupies the dissipative intermediate level and would likewise not realize perfect efficiency (Supplemental Section 4.5).

To emphasize that our protocols retain phase coherence, we utilize them to expedite the transfer and initialization of superposition states (Fig. 4a). Starting with an initial superposition $|\psi_I\rangle = 1/\sqrt{2}\,(|0\rangle + e^{i\phi_I}|-1\rangle)$ and applying STIRAP on the $|-1\rangle$ component, we propagate the initialized phase to the ideal transferred state $|\psi_F\rangle = 1/\sqrt{2}\,(|0\rangle + e^{i\phi_F}|+1\rangle)$. Incoherent effects, such the spontaneous emission, dephasing, and energy uncertainty of $|A_2\rangle$, will decohere the

transferred phase, but can nevertheless result in population transfer. In Fig. 4b, we display the quadrature amplitudes $X$ and $Y$ of $|\psi_F\rangle$ on a polar plot to visualize $\phi_F$ tracking the increment of $\phi_I$ for $\Omega = 2\pi \cdot 133$ MHz. The MOD-SATD and SATD pulses achieve higher phase visibilities $\sqrt{X^2 + Y^2}$ than the adiabatic pulse does, affirming their superiority for coherent manipulations. Moreover, comparing the phase visibility to the square root of the protocol's population transfer efficiency (delineated by the corresponding solid arc in Fig. 4b) reveals that the superadiabatic population transfers are predominantly coherent, while incoherent contributions account for a larger fraction of the adiabatic transfer (Supplementary Section 4.6). Finally, as detailed in Supplementary Section 1, our analytical framework can be extended to derive pulse shapes that accelerate fractional STIRAP (f-STIRAP)[1]. In normal f-STIRAP, the Stokes and pump pulses adiabatically turn-off with a fixed amplitude and phase relation to initialize arbitrary superpositions of the initial and target states. In Fig. 4c, we show that the preparation of $|\psi_F\rangle = 1/\sqrt{2}\,(|-1\rangle \pm |+1\rangle)$ by f-STIRAP achieves an average fidelity of $\mathcal{F} = .93 \pm .01$ for the SATD protocol, an improvement over $\mathcal{F} = .83 \pm .01$ for the adiabatic pulse at $\Omega = 2\pi \cdot 135$ MHz.

Our work establishes SATD as a fast and robust technique for coherent quantum control, with applications to other adiabatic protocols and physical systems. The extension of adiabatic techniques to more open quantum systems highlights the importance of STA as a means to outpace decoherence, without sacrificing robustness. For STIRAP in engineered, solid-state systems involving ladder energy structures[13,14] or cavity-qubit states[16], dissipation is unavoidable as it affects multiple levels, rather than only the intermediate level. In these cases, our 'speed above all' approach with SATD and its flexibility to design transitionless evolutions tailored to specific criteria offer unique advantages. Promisingly, while SATD protocols should adjust for dissipative dynamics, we show that they also possess robust effectiveness. Looking forward, the dissipative $\Lambda$ configuration here is exemplified in a future quantum transducer, where a lossy mechanical mode connects qubits to photons in a quantum network.

**Methods**

*Experimental Sample and Setup*

The experiments here are performed on a naturally-occurring NV center in an electronic grade diamond substrate (Element Six). Characterization of the NV center's properties is presented in the Supplementary Information. A home-built confocal microscopy setup interfaces with a closed-cycle cryostat (Montana Instruments) that holds the sample at 5.5 K. Applying a magnetic field of 252.5 G along the NV axis splits the $|m_s = -1\rangle$ and $|m_s = +1\rangle$ ground states by 1.414 GHz. A 532 nm laser initializes the NV center spin into the $|m_s = 0\rangle$ state with a polarization >90%. Additionally, two tunable 637 nm lasers are actively stabilized on resonance with the $|-1\rangle \rightarrow |A_2\rangle$ and $|0\rangle \rightarrow |E_Y\rangle$ transitions for protocol interaction and spin readout, respectively. The laser resonant with $|-1\rangle \rightarrow |A_2\rangle$ passes through a phase electro-optic modulator (PEOM) to allow simultaneous addressing of $|+1\rangle \rightarrow |A_2\rangle$ by the red-shifted first harmonic. Subsequently, it passes through an amplitude electro-optic modulator (AEOM) to allow sub-nanosecond analog modulation of the intensity. One channel of a 10-GHz clock-speed arbitrary waveform generator (AWG) controls the quadrature modulation of a signal generator (SG) output at the harmonic frequency (1.414 GHz) that is applied to the PEOM, while a second channel directly drives the AEOM. Additional microwave tones at 2.171 GHz and 3.585 GHz are directed to a coplanar waveguide on the sample to manipulate the NV center within its ground state manifold. A second synchronized 1 GHz clock-speed AWG controls the timing of the various tomographic microwave pulses, acoustic-optic modulators, and photon-counting gates. Detailed descriptions of the hardware calibration and the conversion of the Stokes $\Omega_S(t)$ and pump $\Omega_P(t)$ pulse amplitudes to AWG waveforms are described in the Supplementary Information.

*Data Analysis*

After 532 nm excitation and microwave transfer, the initialized state to begin STIRAP is estimated to contain 0.03/0.91/0.06 ($\pm 0.02$) populations in the $|0\rangle/|-1\rangle/|+1\rangle$ states, respectively. To account for this imperfect initialization, we define the transfer efficiency *E*:

$$E \equiv \frac{p_{+1,\text{final}} - p_{+1,\text{initial}}}{p_{-1,\text{initial}} - p_{+1,\text{initial}}} \quad (1)$$

where $p_i$ denotes population of state $|i\rangle$ in the initialized or transferred final state (additional discussion in Supplementary Section 4.1). This simple definition does not properly distinguish the contributions due to incoherent processes, but we verify that this inaccuracy mainly affects the Vitanov shape in the non-adiabatic $A \rightarrow 1$ regime and does not fundamentally alter the conclusions of the paper (Supplementary Section 4.6). To gauge the contribution from coherent transfer, we visualize in Fig. 4b the phase $\phi_F$ of the transferred state $|\psi_F\rangle$ by defining the X and Y amplitudes as the component of the projections on the $1/\sqrt{2}\,(|0\rangle + |+1\rangle)$ and $1/\sqrt{2}\,(|0\rangle + i|+1\rangle)$ basis states that varies the initialized phase $\phi_I$. In general, the mean of these projections will also change due to imperfect transfer into the $|0\rangle\,/\,|+1\rangle$ subspace. Due to different microwave paths to our PEOM, which controls the relative phase of the STIRAP fields[12], and to our coplanar waveguide, which controls the phase of our tomography pulses, we expect in general $\phi_F = \phi_I + \phi_0$. However, we can negate the constant offset $\phi_0$ by appropriate definition of the final state projection basis.

For measurements of fractional STIRAP, we define the density matrix of the final superposition state as

$$\rho = \frac{1}{2}(S_I \hat{\sigma}_I + S_X \hat{\sigma}_X + S_Y \hat{\sigma}_Y + S_Z \hat{\sigma}_Z) \quad (2)$$

where $\hat{\sigma}_I$ is the identity matrix, $\hat{\sigma}_{X/Y/Z}$ are the standard Pauli matrices, and $S_i$ are the corresponding real coefficients, which are plotted in Fig. 4c. The fidelity $\mathcal{F}$ is then computed

from $\rho$ via

$$\mathcal{F} = \left(\text{tr}\left(\sqrt{\sqrt{\rho}\sigma\sqrt{\rho}}\right)\right)^2 = \langle\Psi_{target}|\rho|\Psi_{target}\rangle \tag{3}$$

where $\sigma = |\Psi_{target}\rangle\langle\Psi_{target}|$ is the density matrix of the ideal target superposition.

*Superadiabatic Protocols*

To speed up adiabatic protocols, our general approach aims to enforce transitionless evolution in an arbitrary basis, rather than strictly in the adiabatic basis used by conventional counterdiabatic driving. The dressed dark state defined by this basis choice coincides with the initial and target states at the extremities of the protocol, enabling replication of the result of the adiabatic protocol for arbitrarily short evolution time, but via an alternate trajectory that utilizes all three states of the $\Lambda$ system for STIRAP. Briefly, we note that our generalized approach is fundamentally different from a recent demonstration of counterdiabatic STIRAP in ensembles of rubidium atoms[33]. There, the special condition of large $\Delta$ is needed to adiabatically eliminate the intermediate state from the dynamics and allow established counterdiabatic techniques for two-level systems[17,18] to be employed.

Ref. 10 and Supplementary Section 1 detail analytical derivations for two basis choices: SATD and MOD-SATD, where the latter is derived from the former by reducing the intermediate state occupation. Here, we only state final results critical to understanding the experimental implementation. Starting from the rotating frame Hamiltonian in the basis of the NV center states $\{|0\rangle, |-1\rangle, |+1\rangle, |A_2\rangle\}$,

$$H = \frac{\hbar}{2}\begin{pmatrix} 0 & 0 & 0 & 0 \\ 0 & 0 & 0 & \Omega_P(t) \\ 0 & 0 & 2\delta & \Omega_S(t)e^{i\phi_S(t)} \\ 0 & \Omega_P(t) & \Omega_S(t)e^{-i\phi_S(t)} & 2\Delta \end{pmatrix} \tag{4}$$

the base adiabatic pulse for STIRAP is the 'Vitanov' shape[23]:

$$\Omega_S(t) = \Omega \cos(\theta(t))$$

$$\Omega_P(t) = \Omega \sin(\theta(t)) \quad (5)$$

$$\theta(t) = \frac{\pi}{2} \frac{1}{1 + e^{-\nu(t - L(\nu,\epsilon)/2)}}$$

where $\Omega$ is the angular frequency of Rabi oscillations and the parameter $\nu$ determines the rate of the adiabatic sweep. This shape is considered optimal in the adiabatic limit as it maintains constant $\sqrt{\Omega_S(t)^2 + \Omega_P(t)^2} = \Omega$ over the protocol. Our definition of $\theta(t)$ anticipates the experimental practicality that the $\Omega_S(t)$ and $\Omega_P(t)$ pulses must be truncated in finite time. We define the pulse as existing over $t \in [0, L]$ such that $\Omega_P(0) = \Omega_S(L) = \epsilon \cdot \Omega$. This implies

$$L(\nu, \epsilon) = \frac{2}{\nu} \log\left(\frac{\pi}{2} \frac{1}{\sin^{-1}(\epsilon)} - 1\right). \quad (6)$$

The SATD corrected pulses are given by

$$\Omega_S^{SATD}(t) = \Omega \cos(\theta(t)) - \frac{4 \Omega \sin(\theta(t)) \ddot{\theta}(t)}{\Omega^2 + 4 \dot{\theta}(t)^2}$$

$$\Omega_P^{SATD}(t) = \Omega \sin(\theta(t)) + \frac{4 \Omega \cos(\theta(t)) \ddot{\theta}(t)}{\Omega^2 + 4 \dot{\theta}(t)^2} \quad (7)$$

where we define $t$ on the same interval $[0, L]$. See Supplementary Information for numerical results for MOD-SATD pulses. In order for $\Omega_{S,P}^{SATD}(t)$ to not exceed the original $\Omega$ of the Vitanov pulse, the smallest $\Omega$ that can be corrected is

$$\Omega_{min} = \frac{\nu}{1.315}, \quad (8)$$

using which we define an adiabaticity parameter $\Omega/\Omega_{min}$. When this parameter is applied to the theoretical calculation of pulses, we denote it as the shape parameter $A_{shape} = \Omega_{shape}/\Omega_{min}$, where $\Omega_{shape}$ is the assumed Rabi coupling. When it describes actual experimental conditions, it is distinguished as the experimental adiabaticity $A = \Omega/\Omega_{min}$, where $\Omega$ is the actual experimental coupling. From Eq. 6, we see that, equivalently, the pulse length $L$ defines $\Omega_{min}$ via

$$\Omega_{min} = \frac{1.521}{L(\nu,\epsilon)} \log(\frac{\pi}{2} \frac{1}{\sin^{-1}(\epsilon)} - 1). \tag{9}$$

In the experiment, we set $\epsilon = 10^{-2}$. In Fig. 2 and 4b of the main text, we use $\nu = 0.6$ GHz, which yields $\Omega_{min} = 2\pi \cdot 72.6$ MHz and $L = 16.8$ ns. In Fig. 3, $\nu$ is varied to change the pulse duration $L$. In addition, we allow an arbitrary 2 ns at the start and finish for $\Omega_S(t)$ and $\Omega_P(t)$ to turn on and off via their multiplication by an envelope function (Supplementary Section 2.4). For the ratio of the superadiabatic and adiabatic pulse durations in Fig. 3, we neglect this constant 4 ns of additional time, as this duration does not affect the final transfer efficiency.

*Master Equation Modeling*

The master equation in Lindblad form is given by

$$\dot{\rho} = -\frac{i}{\hbar}[H,\rho] + \sum_k \left(L_k \rho L_k^\dagger - \frac{1}{2} L_k^\dagger L_k \rho - \frac{1}{2} \rho L_k^\dagger L_k\right) \tag{10}$$

where $H$ is Hamiltonian in Eq. 4 and $L_k$ denote the Lindblad operators describing dissipative processes. These include the relaxation rates $\Gamma_0 = 9.8$ MHz, $\Gamma_{-1} = 26.9$ MHz, and $\Gamma_{+1} = 53.5$ MHz for decay of the excited $|A_2\rangle$ level into the $|0\rangle$, $|-1\rangle$, and $|+1\rangle$ ground states, respectively. In addition, the excited state dephases at a rate $\Gamma_{orb} = 55.5$ MHz. The relaxation rates are extracted from measurements of the time-resolved optical pumping through $|A_2\rangle$ into the ground states (Supplementary Section 2). The dephasing rate $\Gamma_{orb}$ is determined from a global fit to optical Rabi data comprising both the $|-1\rangle$ and $|+1\rangle$ to $|A_2\rangle$ transitions over a range of optical powers. The ground state dephasing of $T_2^* \sim 6$ $\mu s$ does not play a significant role over the timescales of the experiment here. In addition, we model the spectral diffusion of the excited state by averaging simulations over a Gaussian distribution of the one photon detuning $\Delta$, with standard deviation $\sigma \sim 2\pi \cdot 31$ MHz (in units of energy/$\hbar$) estimated from an independent measurement. See Supplementary Section 3 for detailed discussion.

**Acknowledgements**

We thank C. F. de las Casas, D. J. Christle for experimental suggestions and P. V. Klimov, G. Wolfowicz for thoughtful readings of the manuscript. C.G.Y., F.J.H., and D.D.A. were supported by the U.S. Department of Energy, Office of Science, Office of Basic Energy Sciences, Materials Sciences and Engineering Division. B.B.Z. and P.C.J. were supported by the Air Force Office of Scientific Research and the National Science Foundation DMR-1306300. In addition, A.A. and G.B. acknowledge support from the German Research Foundation (SFB 767). A.B., H.R., and A.A.C. acknowledge support from the Air Force Office of Scientific Research.

**Author Contributions**

H.R. and B.B.Z. engaged in preliminary discussions. A.B., H.R., and A.A.C. developed the superadiabatic theory. B.B.Z., C.G.Y., F.J.H., and P.C.J. performed the experiments. A.B., A.A., H.R., and G.B. completed the master equation modeling. D.D.A. advised all efforts. All authors contributed to the data analysis and writing of the manuscript.

# Figures

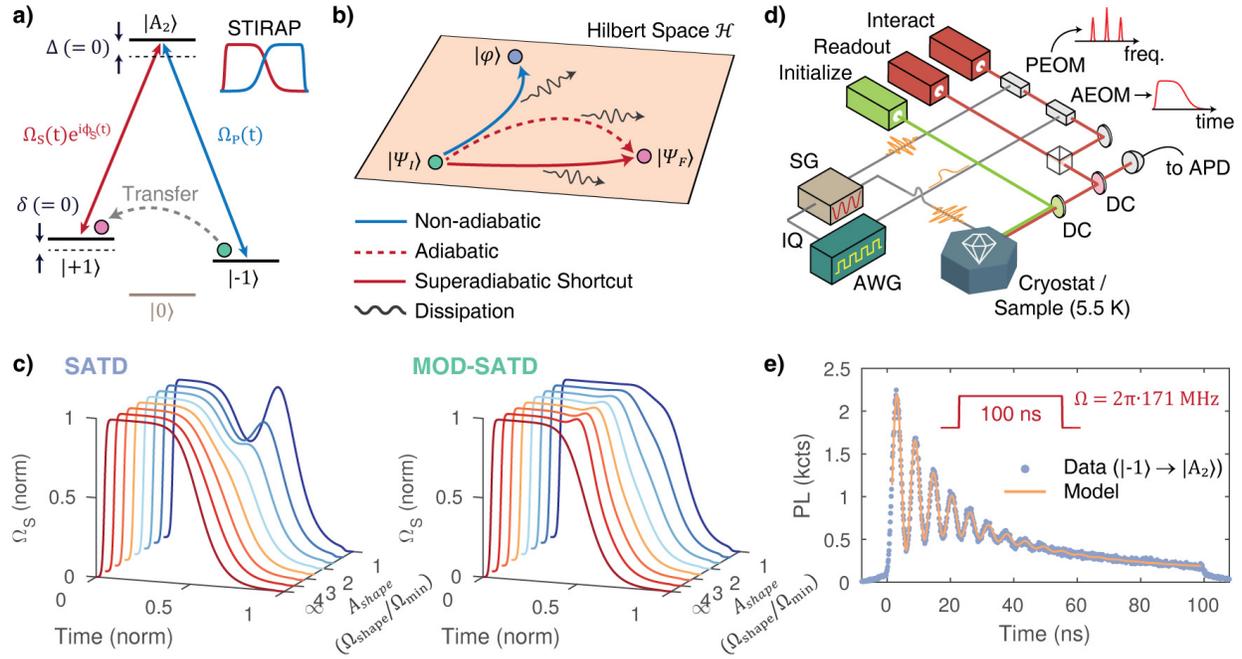

**Figure 1 | Concept and implementation of three-level superadiabatic transitionless driving.**

a) State transfer in a NV center $\Lambda$ system by STIRAP. The $|+1\rangle$ / $|-1\rangle$ ground-state spin levels are coupled by resonant Stokes $\Omega_S(t)$ and pump $\Omega_P(t)$ optical fields to the $|A_2\rangle$ excited state, which acts as the intermediate state for STIRAP. b) Schematic of possible dynamics. An adiabatic protocol transfers the initial state $|\psi_I\rangle$ to the final state $|\psi_F\rangle$ along the dashed, red trajectory, which is followed exactly only in the infinite time limit. For finite-time realizations, $|\psi_I\rangle$ may be transferred to a different state $|\varphi\rangle$ due to non-adiabatic transitions (blue). Our superadiabatic shortcut (solid red) implements modified driving pulses to reproduce the same final transfer of the adiabatic protocol, but for arbitrary evolution time and along a different path determined by the choice of dressed basis. Dissipation leads to errors for all evolutions. c) Example of the modified $\Omega_S(t)$ pulses for SATD and MOD-SATD, corresponding to two different basis choices. The shape parameter $A_{shape}$ specifies the appropriate driving pulse under unitary evolution for a particular experimental coupling strength $\Omega$ and pulse duration $L$. The modified $\Omega_P(t)$ pulses (not shown) mirror the $\Omega_S(t)$ pulses about the midpoint of the protocol. d) Experimental setup utilizing EOMs to shape the Stokes and pump pulses from a single laser on sub-nanosecond timescales. AWG, arbitrary waveform generator; IQ, quadrature modulation; SG, signal generator; P/AEOM, phase/amplitude electro-optic modulator; DC, dichroic mirror; APD, avalanche photodiode. e) Optically-driven Rabi oscillations between the $|-1\rangle$ and $|A_2\rangle$ levels. The oscillations damp due to excited state dissipation (lifetime and dephasing) and spectral diffusion. The solid line is an example of a fit to a master equation model using the rates given in the main text.

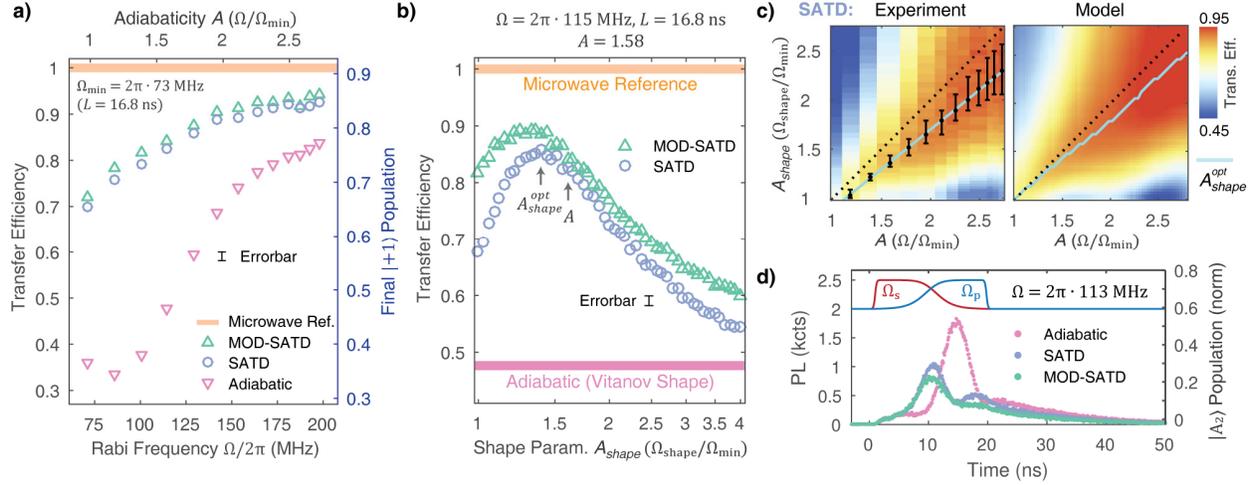

**Figure 2 | Performance and robustness of superadiabatic pulses.**

a) STIRAP transfer efficiency of MOD-SATD, SATD, and adiabatic (Vitanov) pulses as a function of the maximum optical Rabi strength $\Omega$. The superadiabatic protocols utilize pulses prescribed for unitary evolution: the shape parameter $A_{shape}$ is equal to the experimental adiabaticity $A$, determined by $\Omega$ and the constant pulse duration $L$ = 16.8 ns. The right y-axis indicates the absolute population in $|+1\rangle$ at the end of the protocol. The left y-axis estimates a transfer efficiency that accounts for imperfect initialization by using direct microwave transfer from $|0\rangle$ into $|+1\rangle$ to establish a reference transfer efficiency of 1. b) Robustness of the transfer efficiency as a function of the pulse shape $A_{shape}$ for $\Omega = 2\pi \cdot 115$ MHz. The optimal transfer efficiency for the superadiabatic protocols occurs for a shape parameter $A_{shape}^{opt} < A$, reflecting the presence of dissipation and spectral diffusion. Typical errorbars in a) and b) correspond to 95% confidence. c) False color plot of the experimental (left) and simulation (right) transfer efficiency for SATD as a function of $A$ and $A_{shape}$. The dashed black lines represent $A_{shape} = A$. The data points and fitted cyan line on the experimental plot delineate the extracted $A_{shape}^{opt}$, while the interval corresponds to $\pm 1\%$ in transfer efficiency. The deviation $A_{shape}^{opt} < A$ is consistent with the dissipative model (cyan trace denotes $A_{shape}^{opt}$ in model results). d) Photoluminescence (PL) (left y-axis) and converted $|A_2\rangle$ population (right y-axis) measured during the adiabatic, SATD, and MOD-SATD pulses for $\Omega = 2\pi \cdot 113$ MHz, highlighting the designed occupation of $|A_2\rangle$ (less for MOD-SATD) by the superadiabatic pulses.

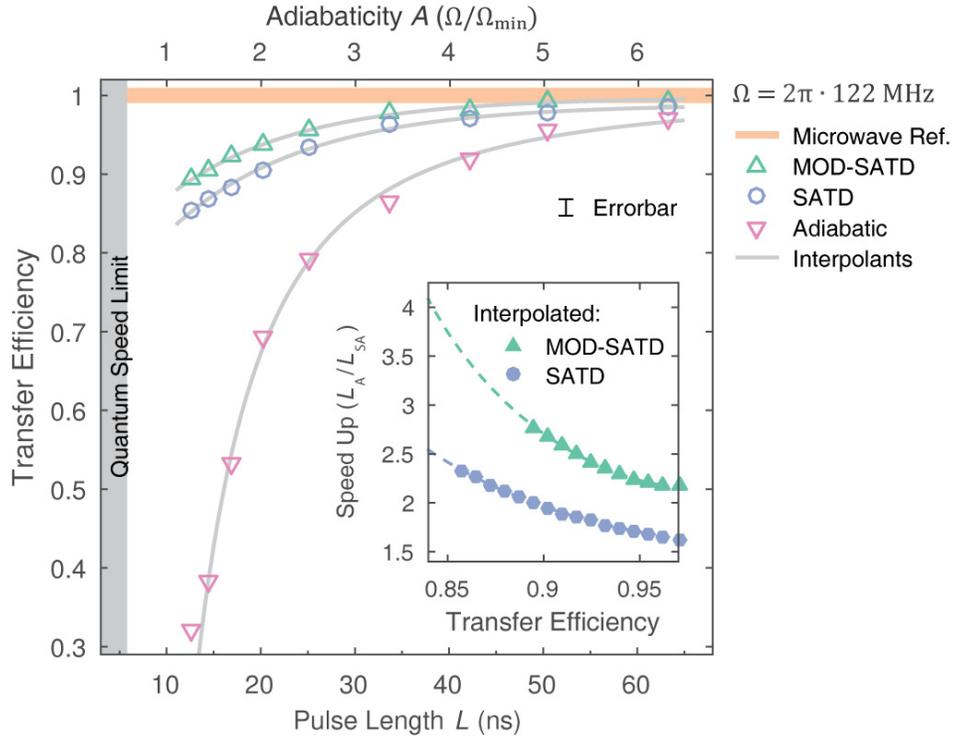

**Figure 3 | Speed-up of superadiabatic protocols.**

STIRAP transfer efficiency for the optimal MOD-SATD and SATD pulses versus the adiabatic pulse as a function of the pulse duration $L \propto \Omega_{min}^{-1}$ for a constant Rabi strength $\Omega = 2\pi \cdot 122$ MHz. The vertical grey bar at 5.8 ns represents the quantum speed limit for state transfer via an intermediate state for this coupling strength $\Omega$. The solid grey lines represent interpolating functions used to invert the plot and estimate the pulse length $L_A$ ($L_{SA}$) of the adiabatic (superadiabatic) protocol needed to attain a given transfer efficiency. The inset displays the speed-up factor, given by the ratio $L_A/L_{SA}$, as a function of the desired transfer efficiency. Dashed lines in the inset represent extrapolations outside the range of experimentally attained transfer efficiencies.

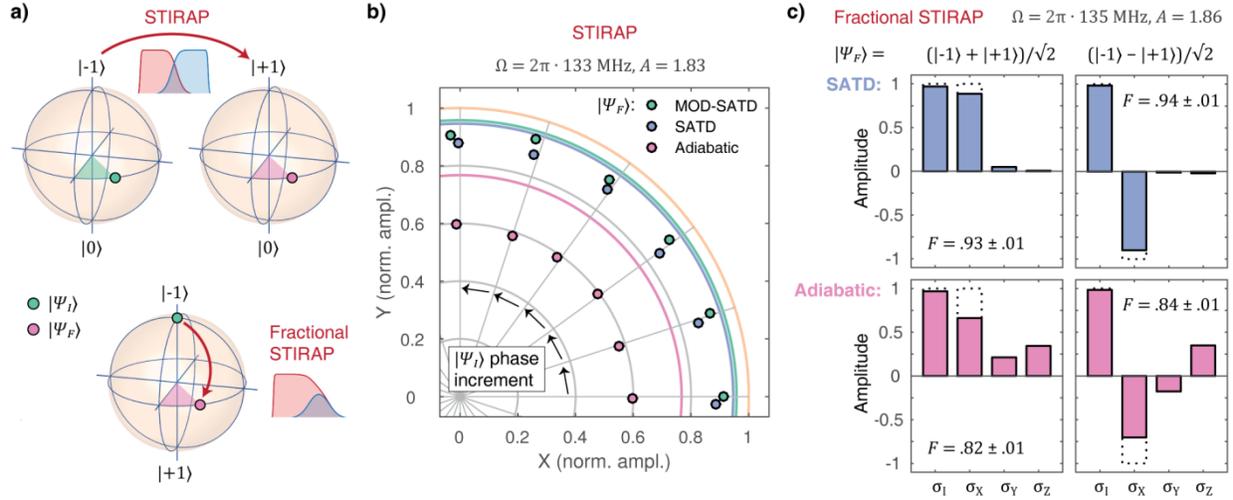

**Figure 4 | Accelerating the transfer and initialization of superposition states.**

a) Bloch sphere schematic for phase-coherent STIRAP processes. (Top) Transfer of superpositions: the phase relation within an initial superposition $|\psi_I\rangle$ of the $|0\rangle$ / $|-1\rangle$ states is transferred by STIRAP to a target superposition $|\psi_F\rangle$ of the $|0\rangle$ / $|+1\rangle$ states. (Bottom) Initialization of superpositions: fractional STIRAP enables the creation of arbitrary superpositions of the $|-1\rangle$ / $|+1\rangle$ states by maintaining a particular phase and amplitude relation between $\Omega_S(t)$ and $\Omega_P(t)$ as both fields are simultaneously ramped to zero. b) Visualization of the phase of the transferred superposition $|\psi_F\rangle$ on a polar plot for MOD-SATD, SATD, and adiabatic protocols as the phase of $|\psi_I\rangle$ is incremented. *X* and *Y* are the components of the projections of $|\psi_F\rangle$ onto $1/\sqrt{2}\,(|0\rangle+|+1\rangle)$ and $1/\sqrt{2}\,(|0\rangle+i|+1\rangle)$, respectively, that vary with the initialized phase. The phase visibility $\sqrt{X^2+Y^2}$ can be compared to the square root of the population transfer efficiency (delineated by the solid arcs) to gauge the coherent fraction of the population transfer for each protocol. c) State tomography and fidelity $\mathcal{F}$ for the initialization of two different final superposition states $|\psi_F\rangle = 1/\sqrt{2}\,(|-1\rangle \pm |+1\rangle)$ by fractional STIRAP via a shortcut SATD protocol (top) and an adiabatic protocol (bottom).